\documentclass[preprint,showpacs,preprintnumbers,amsmath,amssymb,nofootinbib]{revtex4}
\usepackage{bbm}
\usepackage{amsfonts}
\usepackage{booktabs}
\usepackage{mathrsfs}
\usepackage{epsfig}
\usepackage{graphicx}
\usepackage{dcolumn}
\usepackage{bm}
\usepackage{amsmath}
\usepackage{slashed}
\usepackage{subfigure}

\let\jnfont=\rm
\def\NPB#1,{{\jnfont Nucl.\ Phys.\ B }{\bf #1},}
\def\PLB#1,{{\jnfont Phys.\ Lett.\ B }{\bf #1},}
\def\EPJC#1,{{\jnfont Eur.\ Phys.\ Jour.\ C }{\bf #1},}
\def\PRD#1,{{\jnfont Phys.\ Rev.\ D }{\bf #1},}
\def\PRL#1,{{\jnfont Phys.\ Rev.\ Lett.\ }{\bf #1},}
\def\MPLA#1,{{\jnfont Mod.\ Phys.\ Lett.\ A }{\bf #1},}
\def\JPG#1,{{\jnfont J.\ Phys.\ G}{\bf #1},}
\def\CTP#1,{{\jnfont Commun.\ Theor.\ Phys.\ }{\bf #1},}
\def\ZPC#1,{{\jnfont Z.\ Phys.\ C }{\bf #1},}
\def\JHEP#1,{{\jnfont JHEP \ }{\bf #1},}
\def\Rv{\not{\hbox{\kern-1pt $R$}}}
\def\p{\not{\hbox{\kern-3pt $p$}}}

\def\s{s_\beta}
\def\c{c_\beta}
\def\r{x_t}

\begin{document}

\title{Constraining Top partner and Naturalness in the Littlest Higgs model at the LHC and TLEP}

\author{Chengcheng Han$^{1,2}$, Archil Kobakhidze$^4$, Ning Liu$^{1}$, Lei Wu$^4$, Bingfang Yang$^{1}$}
\affiliation{$^1$ Institute of Theoretical Physics, Henan Normal University, Xinxiang 453007, China\\
$^2$ Asia Pacific Center for Theoretical Physics, San 31, Hyoja-dong, Nam-gu, Pohang 790-784, Republic of Korea\\
$^3$ ARC Centre of Excellence for Particle Physics at the Terascale, School of Physics, The University of Sydney, NSW 2006, Australia\\
}%

\date{\today}

\begin{abstract}

We investigate indirect constraints on the top partner in the Littlest Higgs model. By performing a global fit of the latest Higgs data,  $B_s \to \mu^+\mu^-$ measurements and the electroweak precision observables we find that the top partner with the mass up to 830 GeV is excluded at $2\sigma$ level. Our bound on the top partner mass is much stronger than the bounds obtained from the direct searches at the LHC. Under the current constraints  the fine-tuning measure is less than 9\% and the branching ratio of $T \to tZ$ is bounded between 14\% and 25\%. We also find that precise measurements of Higgs couplings at 240 GeV TLEP will constrain the top partner mass in multi-TeV region.

\end{abstract}

\maketitle


\section{\label{sec:level1}Introduction}

The measured properties of the recently discovered 125 GeV Higgs boson  \cite{higgs-atlas,higgs-cms} are in a very good agreement with the Standard Model (SM) predictions. The experimental errors, however, are still large enough and to various deviations from the SM can still be accommodated. In fact, theoretical considerations on the radiative stability of the Higgs boson mass are widely considered as a major motivation for new physics beyond the SM, which ameliorates the fine-tuning between the bare Higgs mass and the quadratically divergent radiative correction. The radiative stability of the Higgs mass is typically attributed to a new symmetry such as softly-broken supersymmetry \cite{susy} or spontaneously broken global symmetry as in the Little Higgs models \cite{littlehiggs}.  These extensions of the SM predict new particles which contribute to the radiatively corrected Higgs mass, cancelling quadratically divergent contributions from the SM particles, most notably,  the top quark contribution. Within the supersymmetric models this role is played by the sub-TeV spin-0 top partner, the top squark, while in Little Higgs models the top partner is a spin-1/2 vector-like quark. The search for top partners, therefore, is an important task, as it may shed light on the long standing naturalness problem \cite{stop,eff-ftop,com-ftop,lh-ftop,tp-model}.

Compared with the scalar top partners, the fermionic top partner has larger production rate and simpler decay modes at colliders than the scalar top partner of the same mass. Constraints from the direct fermionic top partner pair production searches have been presented by the ATLAS and CMS at 7+8 TeV LHC. The bound on the top partner mass is sensitive to branching ratios of the top partner decays into different final states $bW$, $tZ$ and $th$. The top partner with the mass less than $687-782$ GeV were shown to be excluded \cite{tp-atlas,tp-cms}. Different strategies have been suggested to improve the discovery sensitivity of the top partners. For improved analysis of top partner pair production processes the use of the jet substructure technique were proposed in Ref. \cite{tp-jet}. If the top partner mass in the range of $600-1000$ GeV, single top partner production can have larger cross section than the pair production and, hence, it may be more favourable to look for singly produced top partners at the LHC \cite{tp-singly}.

In addition to direct searches one can exploit indirect searches for the top partners through their contribution to the electroweak precision observables \cite{tp-ewpo} and flavor physics \cite{tp-flavor}. Also, since top partner is naturally related to the Higgs physics,
one can obtain constraints from the Higgs data \cite{tp-higgs}. The indirect searches become increasingly important for heavy top partners, which may not be directly observable at the LHC. The study of indirect effects of top partners are of great importance for future colliders as well.

In this work, we will study a simplified fermionic top partner model, which can be considered as the top sector of the Littlest Higgs (LH) model \cite{lh}. There are many phenomenological works devoted to study it before the discovery of the Higgs boson. We perform a state-of-the-art global fit to obtain the indirect constraints on fermionic top partner with a comprehensive way. This method was widely used in the fit of the SM to the electroweak precision data and has been recently used in the studies of the parameters space of the supersymmetric models, such as cMSSM, pMSSM and NMSSM. So, it will be also meaningful to explore what might happen in a fermionic top partner model with a global fit at future colliders. Our study may play a complementary role to the direct searches in probing top partner. More importantly, by building an overall likelihood function for the constraints from the Higgs data, $B_s \to \mu^+\mu^-$ measurements and the electroweak precision observables, we can obtain a well-defined statistical results of the exclusion limit on the top partner. On the other hand, we explore the potential of constraining the top partner from the future Higgs couplings measurements at TLEP. The proposed TLEP $e^+e^-$ collider \cite{tlep} could be located in a new 80 to 100 km tunnel in the Geneva area. It would be able to produce collisions at 4 interaction points with $\sqrt{s}$ from 90 to 350 GeV and beyond and is expected to make precision measurements at the $Z$ pole, at the $WW$ threshold, at the $HZ$ cross section maximum, and at the $t\bar{t}$ threshold, with an unprecedented accuracy. The luminosity expected at TLEP is between a factor 5 and 3 orders of magnitude larger than that expected for a linear collider, such as ILC and CLIC. In light of the high luminosity, TLEP can allow to measure the Higgs couplings to percent level, such as $hVV$ and $h\gamma\gamma$, which are sensitive to the possible new physics that can reduce the fine-tuning of the Higgs boson mass.

This paper is organized as follows. In section II, we give a brief description of the simplified fermionic top partner model. In section III, we present the numerical results and discussions. Finally, we draw our conclusions in section IV.

\section{The Model}

The generic structure of the Littlest Higgs models employs a global symmetry broken at a TeV scale, where new particles cancel divergences from the SM particles in the Higgs mass calculation \cite{lh}. An extended gauge and scalar interactions in the full theory contribute to the fine-tuning in a rather complicated and model-dependent way \cite{lh-variants}. The most relevant for the Higgs mass naturalness problem, however, is the top quark sector. Therefore, we can simplify the top quark sector of the Littlest Higgs model based on non-linear realization, where an extra global $SU(3)$ symmetry is dynamically broken down to $SU(2)$ at a scale $f$ by some unspecified strong dynamics. The low-energy non-linear field which spans the coset $SU(3)/SU(2)$ is defined as:
\begin{equation}\label{sigma}
V = \exp \left(\frac{i \pi_a t_a}{f}\right) \left( \begin{tabular}{c} 0\\0\\$f$ \end{tabular} \right) \,,
\end{equation}
where $t_a$ are the broken generators ($a=1\ldots 5$), $\pi_a$ are the corresponding Goldstone bosons. Four of these Goldstone states are combined into the SM electroweak Higgs doublet $H$, while the remaining singlet does not play any role in our analysis and we ignore it in what follows.  The top quark Yukawa coupling is generated by the following interactions \cite{lh,yukawa},
\begin{equation}\label{top-yukawa}
{\cal L} \,=\, -\lambda_1 u_R^\dagger  V^\dagger \chi_L \,-\, \lambda_2 f U_R^\dagger U_L \,+\,~{\rm h.c.}
\end{equation}
where $\chi_L = (\sigma^2 Q, U)^T_L$ is an $SU(3)$ triplet of left-handed Weyl fermions, and $u_R$ and $U_R$ are two $SU(3)$ singlet right-handed Weyl fermions. While the first term in Eq.(\ref{top-yukawa}) is $SU(3)-$symmetric, the second term explicitly violates $SU(3)$ global invariance and, hence, the Higgs mass is generated through the radiative corrections owing to this violating term.

After the electroweak symmetry breaking, the top mass terms are given by,
\begin{equation}
{\cal L}_{\rm mass} = (u_R^\dagger~U_R^\dagger)\, {\cal M} \,\left( \begin{tabular}{c} $u_L$ \\$U_L$ \end{tabular} \right) \,+\,~{\rm h.c.} \,,
\end{equation}
with
\begin{equation}\label{mass}
{\cal M} \,= \,f\,\left( \begin{tabular}{cc} $\lambda_1 \sin \bar{a}$ & $\lambda_1 \cos\bar{a}$ \\ $0$ & $\lambda_2$ \end{tabular} \right)\,.
\end{equation}
where $\bar{a} = v/(\sqrt{2}f)$ and $v$ is the vev of Higgs field. The mass matrix Eq.(\ref{mass}) can be diagonalized by rotating the weak eigenstates $(u,U)$ to the mass eigenstates $(t,T)$,
\begin{eqnarray}
t_L &=& \cos\beta \,u_{L} - \sin\beta \,U_{L},~~~~
T_{L} = \sin\beta \,u_{L} +\cos\beta \,U_{L}\nonumber \\
t_R &=& \cos\alpha \,u_R - \sin\alpha \,U_{R},~~~~
T_{R} = \sin\alpha \,u_R + \cos\alpha \,U_{R}\
\end{eqnarray}
where $t=(t_L,t_R)$ and $T=(T_L,T_R)$ are identified with the SM top quark and the top partner respectively. In the mass egeinstate, the Eq.(\ref{top-yukawa}) becomes
\begin{equation}
{\cal L}_{\rm int} \,=\, -\lambda_t t_R^\dagger  \tilde{H} Q_L \,-\, \lambda_T T_R^\dagger  \tilde{H} Q_L
\,+\, \frac{\lambda_1^2}{m_T} (H^\dagger H) T_R^\dagger T_L \,+\, \frac{\lambda_1\lambda_2}{2m_T} (H^\dagger H) t_R^\dagger T_L +~{\rm h.c.}~+\ldots
\end{equation}
with
\begin{equation}
\lambda_t = \frac{\lambda_1\lambda_2}{\sqrt{\lambda_1^2+\lambda_2^2}}\,,~~~\lambda_T = \frac{\lambda_1^2}{\sqrt{\lambda_1^2+\lambda_2^2}}\,.
\label{a}
\end{equation}
where $\lambda_t$ is the SM top Yukawa coupling and $m_{T}$ is the mass of the top partner.

Note that, due to the known top quark mass, the three free parameters ($\lambda_{1,2}$ and $f$) are reduced to two, which can be chosen as two physical parameters $\alpha$ and $m_T$. While the left-handed mixing angle $\beta$ can be given in terms of $\alpha$ and $m_T$ as $\sin\beta=x^{1/2}_{t}/\sqrt{\cot^2\alpha+x_t}$. It is clear that the potentially dangerous quadratic divergent contribution to the Higgs mass due to the top quark loop coming from the first term in Eq.(\ref{a}) is cancelled by the top partner loops  originated from the next two terms. The dominant negative log-divergent correction to the Higgs mass squared from the top and $T$ loops is given by \cite{lh}
\begin{equation}
	\delta \mu^2 = -\frac{3 \lambda_t^2 m_{T}^2}{8 \pi^2} \log{\frac{\Lambda^2}{m_{T}^2}}.
\end{equation}
where $\Lambda=4 \pi f$ is the UV cut-off of the model. Then the fine-tuning can be quantified by the following parameter:
\begin{equation} \label{finetuning}
	\Delta^{-1}=\frac{\mu_{\text{obs}}^2}{|\delta \mu^2|}, \qquad \mu_{\text{obs}}^2= \frac{m_h^2}{2}.
\end{equation}
Here $m_h$ is the Higgs boson mass.

Since the main focus of this phenomenological model is on the naturalness problem of the Higgs mass, we simply assume that the gauge sector is the same as the one in the SM \footnote{Note, however, that the Higgs-gauge couplings are suppressed by a factor $\cos\bar{a}$ with respect to the SM predictions.}. The low-energy effects of the underlying strongly coupled sector of the full theory is parameterized by dimension-6 operators \cite{zyhan},
\begin{equation}\label{operators}
{\cal L}_{\rm UV} = \frac{c_1}{\Lambda^2} \left( V^\dagger D_\mu V \right)^2 \,+\, \frac{g g^\prime c_2}{\Lambda^2} W^a_{\mu\nu} B^{\mu\nu}  (V^\dagger Q^a V)\,,
\end{equation}
where $c_1$ and $c_2$ are dimensionless couplings and are expected to be of order 1. These interactions contribute to the electroweak scale observables.

\section{Numerical results and discussions}
In our numerical calculations we take the following SM input parameters \cite{pdg}:
\begin{eqnarray*}
m_t &=& 173.5{\rm ~GeV},~~m_{W}=80.385 {\rm ~GeV}, ~~\alpha(m_Z)= 1/127.918,~~\sin^{2}\theta_W=0.231.
\end{eqnarray*}
A light top partner with mass around weak scale is welcomed by the naturalness, however, it has already been excluded by the electroweak precision observables alone. For similar reason, a small or large mixing angle $\alpha$ is also not favored. Besides, a large $\alpha$ can cause a significant deviation of the result of $B_s \to \mu^+\mu^-$ from the SM prediction. So, in our calculations, we require $m_T>500$ GeV and $0.2<\alpha <1.1$ in our scan. Although there are no a explicit values of UV couplings and the variation of these couplings will have some effects on the observables, the natural values of these couplings in the unknown strong interaction sector should be order one. We take $c_1=c_2=1$ for simplicity.

Our global fit is based on the frequentist theory. For a set of observables ${\cal O}_i (i=1...N)$, the experimental measurements are assumed to be Gaussian distributed with the mean value ${\cal O}^{exp}_i$ and error $\sigma^{exp}_i$. The $\chi^2$ can defined as $\chi^2 = \displaystyle{\sum_{i}^{N}}\frac{({\cal O}^{th}_i-{\cal O}^{exp}_i)^2}{{\sigma_i}^2}$, where $\sigma_i$ is the total error with quadric added the experimental and theoretical errors. The likelihood $\cal L$ for a point in the parameter space is calculated by using the $\chi^2$ statistics as a sum of individual contributions from the above listed experimental constraints. The confidence regions are evaluated with the profile-likelihood method from tabulated values of $\delta\chi^2 \equiv -2 \ln({\cal L} / {\cal L}_{max})$. In two dimensions, 68.3\% confidence regions are given by $\delta\chi^2 =2.30$ and 95.0\% confidence regions by $\delta\chi^2 = 5.99$. In our fit, we vary the mixing angle $\alpha$ and top partner mass $m_T$ within the
following ranges,
\begin{eqnarray}
0.2~ \le \alpha \le 1.1~,~~~0.5~{\rm TeV} \le m_T \le 5~{\rm TeV}.
\end{eqnarray}
The likelihood function $\cal{L}\rm{\equiv exp}[{-\sum \chi^2_i}]$ is constructed from the following constraints:
\begin{itemize}
\item[(1)] The electroweak precision observables: $S$, $T$ and $U$. Similar to the Littlest Higgs, firstly, the top partner can correct the propagators of the electroweak gauge bosons at one-loop level, which is given by \cite{stu},
\begin{eqnarray}
S_T &=& \frac{\s^2}{2\pi}\,\left[ \left(\frac{1}{3}- \c^2\right)\,\log \r \,+\,\c^2 \frac{(1+\r)^2}{(1-\r)^2}\,+\,\frac{2\c^2\r^2(3-\r)\log \r}{(1-\r)^3}\,-\frac{8\c^2}{3}
\right] \\ \nonumber
T_T &=& \frac{3}{16\pi}\,\frac{\s^2}{s_w^2c_w^2}\,\frac{m_t^2}{m_Z^2}\,\left[\frac{\s^2}{\r}-1-\c^2-\frac{2\c^2}{1-\r}\log \r \right] \\ \nonumber
U_T &=& -\frac{\s^2}{2\pi}\,\left[ \s^2\,\log \r \,+\,\c^2\frac{(1+\r)^2}{(1-\r)^2}\,+\,\frac{2\c^2\r^2(3-\r)\log \r}{(1-\r)^3}\,-\frac{8\c^2}{3} \right]
\end{eqnarray}
where $\r=m_t^2/m_T^2$, and $\theta_w$ is the Weinberg angle. Secondly, due to the composite nature of the Higgs boson, the $S$ and $T$ parameters are modified by the deviation of the Higgs gauge couplings $hVV$ from the SM prediction, which is given by \cite{stu},
\begin{eqnarray}
S_h &=& - \frac{1}{3\pi} \frac{m_W^2}{g^2f^2}\,\log \frac{m_h}{4\pi f} \\ \nonumber
T_h &=& \frac{3}{4\pi c_w^2} \frac{m_W^2}{g^2f^2}\,\log \frac{m_h}{4\pi f}
\end{eqnarray}
Thirdly, the 6-dimension operators from the strongly coupled sector in Eq.(\ref{operators}) also contribute to the $S$ and $T$ parameters, \cite{zyhan}
\begin{eqnarray}
S_{\rm UV} &=& \frac{4 c_1 m_W^2}{\pi g^2 f^2} \\ \nonumber
T_{\rm UV} &=& - \frac{c_2 m_W^2}{2 \pi e^2 g^2 f^2}
\end{eqnarray}
The experimental values of $S$, $T$ and $U$ are taken from Ref. \cite{pdg}.

\item[(2)] B-physics. Since the SM flavor symmetry is broken by the extension of the top quark sector, the mixing between top partner and down-type quark can induce flavor changing neutral current processes at one-loop level \cite{lh-flavor}. Among them, the most sensitive one is the rare decay $B_s \to \mu^+\mu^-$. At order $(v/f)^2$, the ratio of the branching ratio of $B_s\to\mu^+\mu^-$ with respective to the SM prediction can be written as \cite{lh-bsmumu},
\begin{equation}
\frac{{\rm Br}(B_s\to\mu^+\mu^-)}{{\rm Br}(B_s\to\mu^+\mu^-)_{\rm SM}}\,=\,\left| 1+\frac{\bar{Y}}{Y_{\rm SM}}\right|^2,
\end{equation}
where
\begin{eqnarray}
& &\hskip-.7cm Y_{\rm SM} = \frac{x_t}{8}\left[ \frac{x_t-4}{x_t-1} +\frac{3x_t}{(x_t-1)^2}\log x_t\right], \\ \nonumber
& &\hskip-.7cm \bar{Y} = s_\beta^2 \left[ \frac{2+2x_t-2x_t^2}{8(-1+x_t)} - \frac{x_t(2-x_t+2x_t^2)}{8(-1+x_t)^2}\log x_t +
\frac{3+2x_t}{8}\log x_T + \frac{x_t}{8}\tan^2\alpha \right].
\end{eqnarray}
The latest combined result from the CMS and LHCb measurements has shown $Br_{exp}(B_s \to \mu^+\mu^-)=(2.9 \pm 0.7)\times 10^{-9}$ \cite{bs-exp}, which is well consistent with the SM prediction $Br_{SM}(B_s \to \mu^+ \mu^-)=(3.56\pm0.30) \times 10^{-9}$ \cite{bs-th}.

\item[(3)] Higgs data. The signal strength of one specific analysis from a single Higgs boson can be given by
\begin{equation}
\mu = \sum_{i} c_i\omega_i,
\label{Eq:mu}
\end{equation}
where the sum runs over all channels used in the analysis. For each channel, it is characterized by one specific production and decay mode. The
individual channel signal strength can be calculated by
\begin{equation}
c_i=\frac{\left[\sigma\times BR\right]_i}{\left[\sigma_{SM}\times BR_{SM}\right]_i},
\label{Eq:ci}
\end{equation}
and the SM channel weight is
\begin{equation}
\omega_i=\frac{\epsilon_i\left[\sigma_{SM}\times BR_{SM}\right]_i}{\sum_j\epsilon_j\left[\sigma_{SM}\times BR_{SM}\right]_j}.
\label{Eq:omega}
\end{equation}
where $\epsilon_i$ is the relative experimental efficiencies for each channel. But these are rarely quoted in experimental publications. In this case, all channels considered in the analysis are treated equally, i.e. $\epsilon_i=1$. We confront the modified Higgs-gauge interactions  $hVV$, $hgg$ and $h\gamma\gamma$ within our model with the Higgs data by calculating the $\chi^{2}_{H}$ of the Higgs sector using the public package \textsf{HiggsSignals-1.2.0} \cite{higgssignals}, which includes 81 data sets from the ATLAS, CMS, CDF and D0 collaborations. We choose the mass-centered $\chi^2$ method in the package \textsf{HiggsSignals}.

\end{itemize}

\begin{figure}[ht]
\centering
\includegraphics[width=3.5in,height=3in]{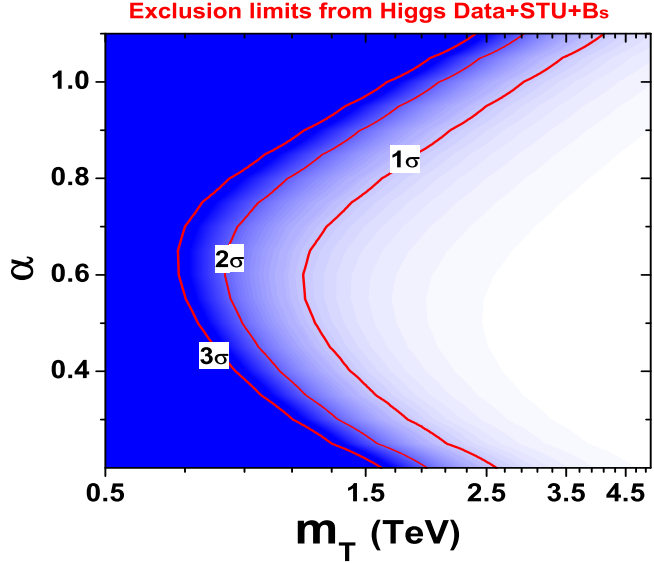}
\vspace{-0.3cm}
\caption{The global fit of the constraints (1)-(3) on the minimal fermionic top partner model in the $\alpha-m_T$ plane. The red lines from right to left respectively correspond to $1\sigma$, $2\sigma$ and $3\sigma$ exclusion limits.}
\label{amt}
\end{figure}

In Fig.\ref{amt}, we show the results of the global fit to the above constraints (1)-(3) in the plane of mixing angle ($\alpha$) versus top partner mass ($m_T$). It can be seen that the combined indirect constraints can exclude the top partner mass up to about 830 GeV at 95\% C.L.. This bound is much stronger than the lower limit set by the ATLAS direct searches for the $SU(2)$ singlet top partner, $m_T > 640$ GeV \cite{tp-atlas}. The allowed low values of $m_T$ are around $\tan\alpha\sim 1$, where top partner contribution to the oblique parameters is minimised. However, it is worth noting that a light top partner with the large mixing angle is strongly disfavoured by the latest result of $B_s \to \mu^+\mu^-$, which provides more stringent bound than the constraint from the oblique parameters. In Ref.\cite{tp-model}, the authors also showed the constraint on the top partner but only from the electroweak precision data. While we present a combined bound on the top partner from the building of an overall likelihood for the electroweak precision observables, Higgs data and $B_s \to \mu^+\mu^-$ measurements. So, the lower bound on the top partner is pushed up from about 500 GeV to 830 GeV.

\begin{figure}[ht]
\centering
\includegraphics[width=4in,height=3.5in]{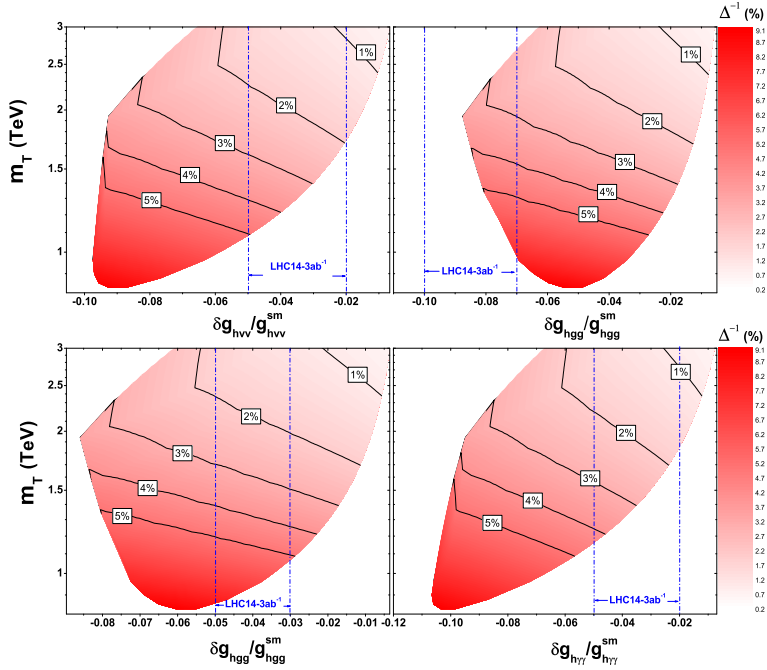}
\vspace{-0.3cm}
\caption{The relative shifts of the Higgs couplings for the samples in the $2\sigma$ allowed range in Fig.\ref{amt}. The dash-dot lines represent the expected measurement uncertainties at HL-LHC \cite{snowmass}.}
\label{couplings}
\end{figure}

In Fig.\ref{couplings}, we show the relative shifts of the Higgs couplings for the above samples in the $2\sigma$ range and compare them with the corresponding expected measurement uncertainties of the Higgs couplings at LHC14 with a luminosity of 3000 $fb^{-1}$ \cite{snowmass}. The fine-tuning for each point is also calculated by using the measure in Eq.(\ref{finetuning}). From Fig.\ref{couplings}, we observe the following: (1) The values of the fine-tuning for the samples are constrained to be smaller than about 9\% by the global fit; (2) Since the Higgs-gauge couplings are suppressed by the common factor $\cos\bar{a}$, the correction to $g_{hVV}$ is always negative in the given model. Also, the loop-induced couplings $g_{h\gamma\gamma}$ and $g_{hgg}$ are reduced due to the cancellation between top quark and the top partner contributions; (3) All the Higgs couplings deviate from the SM predictions at a percent level. Thus, the future measurements of the $g_{h\gamma\gamma}$ coupling at the HL-LHC will be able to exclude the fermionic top partner with $m_T<1.84$ TeV. This corresponds to the fine-tuning being lager than about 2\%. The measurements of $g_{hgg}$ couplings, can only mildly improve the limits on the top partner mass, while measurements of the top quark Yukawa coupling will not provide further constraint on the top partner mass due to the large uncertainties in its determination at the HL-LHC. Note that, in Ref.\cite{tp-model}, the authors calculated the normalized events for $h \to \gamma\gamma$ and $h \to WW$ but without imposing any constraints on the plane of $m_T-\alpha$ from the experimental data. Actually, some regions with fine tuning $\Delta^{-1} >20\%$ have already been excluded by our studied observables. On the contrary, in Fig.2, we require our samples to satisfy the combined constraints at $2\sigma$ level, then calculate the fine tuning for each points in the allowed region. So, we found that the suppression of the couplings of $hVV$ and $h\gamma\gamma$ are at most 10\% and the parameter space with fine tuning $\Delta^{-1}> 10\%$ has been strongly disfavored by the current data.

\begin{figure}[ht]
\centering
\includegraphics[width=3in,height=2.5in]{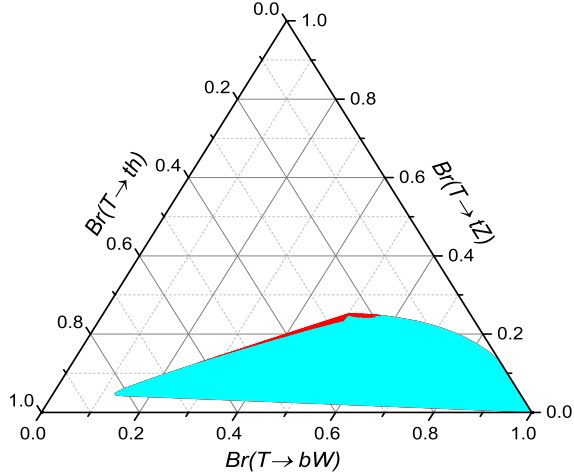}
\caption{The branching ratios of the top partner. The red(cyan) region corresponds to the samples in(out) $2\sigma$ allowed range in Fig.\ref{amt}}
\label{br}
\end{figure}
In Fig.\ref{br} we show the branching ratios of various decays of the top partner. The cyan region is excluded by the global fit at $2\sigma$ level. In the allowed red region, we can find that the branching ratio of $T \to tZ$ is bounded between 14\% and 25\%. The branching ratio for $T \to th$ decay can be competitive to the one of $T \to bW$ for the mixing angle $\alpha \lesssim 0.3$ and can reach up to 57\%. For larger $\alpha$ the branching ratio of $T \to bW$ increases and becomes the main decay mode for $m_T \gg v$. We note that, in Ref.\cite{rao}, the authors have given the strong bounds on the top partner mass ranging from $m_T > 415$ GeV to $m_T > 557$ GeV at 95\% C.L. by combining results of specific $T \to tZ$ and $T \to Wb$ searches, which is across the entire space of branching ratios. As a complementary to this direct bound, our results seem stronger than theirs due to the combination of the recent indirect measurements. So, under the current constraints, $T \to bW$ and $T \to th$ might be the most promising channels for searching for the fermionic top partner at the LHC.

\begin{figure}[ht]
\centering
\includegraphics[width=4in,height=3in]{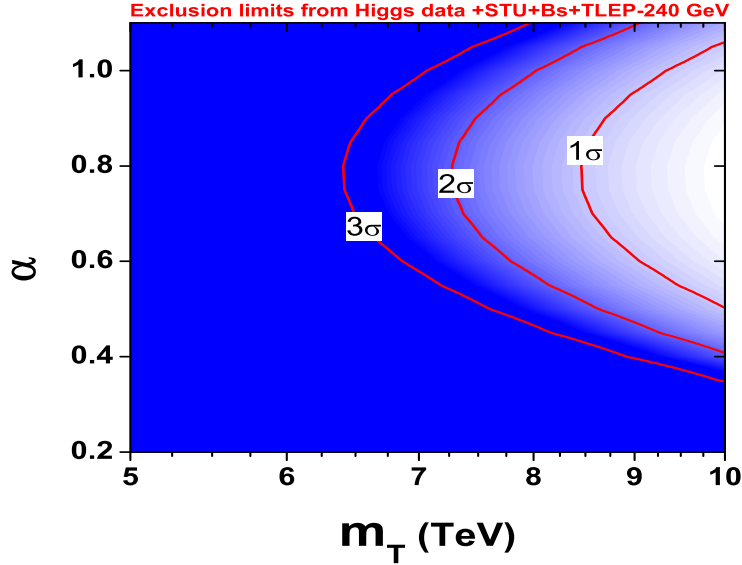}
\caption{The expected exclusion limits on the $m_T -\alpha$ plane from the global fit of the current Higgs data, electroweak observables, $B_s \to \mu^+\mu^-$ and TLEP.}
\label{tlep}
\end{figure}

In Fig.\ref{tlep} we present the prospect of improving the constraints on top partner at discussed future Higgs factory TLEP with $\sqrt{s}=240$ GeV. At $\sqrt{s}=240$ GeV, the TLEP luminosity is expected to be $5 \times 10^{34}$ cm$^{-2}$ s$^{-1}$ at each interaction point, in a configuration with four IPs. So the huge Higgs events allow the Higgs couplings to be measured at percent level at TLEP. By a model-independent fit, expected uncertainties on the measurements of the Higgs gauge couplings $g_{hZZ}$, loop couplings $g_{h\gamma\gamma}$ and $g_{hgg}$ are estimated as 0.16\%, 1.7\% and 1.1\%, respectively at TLEP by the Snowmass Higgs working group. In the fitting, we use the Snowmass Higgs working group results to simply estimate the exclusion limits \cite{snowmass}. We assume that all the measured Higgs couplings will be the same as the SM couplings with the expected measurement uncertainties given in Table 1-16 of Ref. \cite{snowmass} for super-high TLEP luminosities. From the Fig.\ref{tlep} we can see that the lower limit of the fermionic top partner mass will be pushed up to 7.25 TeV and the mixing angle $\alpha$ can be limited to be larger than 0.4 at 95\% C.L..

\section{conclusion}

In this paper, we investigated a minimal $SU(2)$ singlet fermionic top partner model using the available data from the LHC and the electroweak precision observables. By performing the global fit, we find that the top partner mass can be excluded up to 830 GeV at $2\sigma$ level, which is much stronger than the
results of direct searches given by the ATLAS and CMS collaborations. The precise measurements of the Higgs couplings at the future collider, such as TLEP, will improve this limit up to about 7.25 TeV.

\section*{Acknowledgement}
Chengcheng Han was supported by a visitor program of Henan Normal University, during which part of this work was finished. This work was supported by the Australian Research Council, by the National Natural Science Foundation of China (NNSFC) under grants No. 11222548, 11275057 and 11305049, by Specialized Research Fund for the Doctoral Program of Higher Education under Grant No.20134104120002 and by the Startup Foundation for Doctors of Henan Normal University under contract No. 11112.

\end{document}